# Adapting the 23m LST-North mechanical structure design for the strong Chile seismic environment

T. Schweizer*[a], J. Eder[a], Teshima Masahiro[a], Holger Wetteskind[a],
Razmik Mirzoyan[a],
[a]Max Planck Institut für Physik, Boltzmannstr. 8, 85748 Garching, Germany

## ABSTRACT

The 23-meter-diameter Large-Sized Telescope (LST) is the largest size telescope of the next generation Cherenkov Telescope Array (CTA). The first telescope, LST-1, was installed at the Roque de los Muchachos Observatory (ORM) on La Palma at an altitude of 2,250 m in 2018 and has been in operation since 2019. Its ultra-lightweight structure (110 tons) enables extremely rapid repositioning (180° in 18 seconds) and has been designed to withstand extreme environmental conditions, including storms and winds exceeding speed up to 200 km/h. To deploy the same solid telescope design at the CTA Southern Observatory in Chile, the structure must be adapted to the significantly higher seismic demands of the site. To address this challenge, the Max Planck Institute for Physics (MPP) has proposed the integration of a seismic isolation system with the proven LST-1 structural design. This approach substantially reduces seismic loads and dynamic amplification, thereby avoiding extensive structural modifications and enabling the existing telescope design to be transferred to the Chilean site with only minor adaptations. In this contribution, we present the proposed seismic isolation concept and its feasibility studies. MPP is responsible for the mechanical structure of the LST and has validated the concept through detailed finite-element analyses and long-term structural lifetime assessments.

## 1. INTRODUCTION

The 23-meter-diameter Large-Sized Telescope (LST) of the Cherenkov Telescope Array (CTA) was developed over an approximately eight-year design phase between 2008 and 2016. The first telescope, LST-1, was successfully constructed and commissioned at the Observatorio del Roque de los Muchachos (ORM) on the Canary Island of La Palma at an altitude of 2,200 m above sea level. Since 2019, it has been in continuous operation and, as of 2026, regularly acquires scientific data alongside the neighbouring MAGIC telescopes.

The outstanding performance of the LST design has led to its replication, and three additional telescopes (LST-2 to LST-4) are currently undergoing installation and commissioning at the same site. Together, the four telescopes will form the world's most sensitive Cherenkov telescope array in the low-energy gamma-ray regime, achieving an unprecedented energy threshold of down to 20 GeV. During the commissioning and operation of LST-1, valuable experience was gained, leading to the resolution of several technical challenges and the implementation of numerous improvements. The mechanical design of the LST has proven to be robust and solid, making it an attractive candidate for future Cherenkov telescope installations at other sites. From the outset, the telescope structure was designed to withstand severe environmental conditions, including extreme wind loads and significant seismic events.

*tschweiz@mpp.mpg.de; phone +49 89 32354 227; http://mpp.mpg.de,
+josef.eder@t-online.de; www.e-der.de

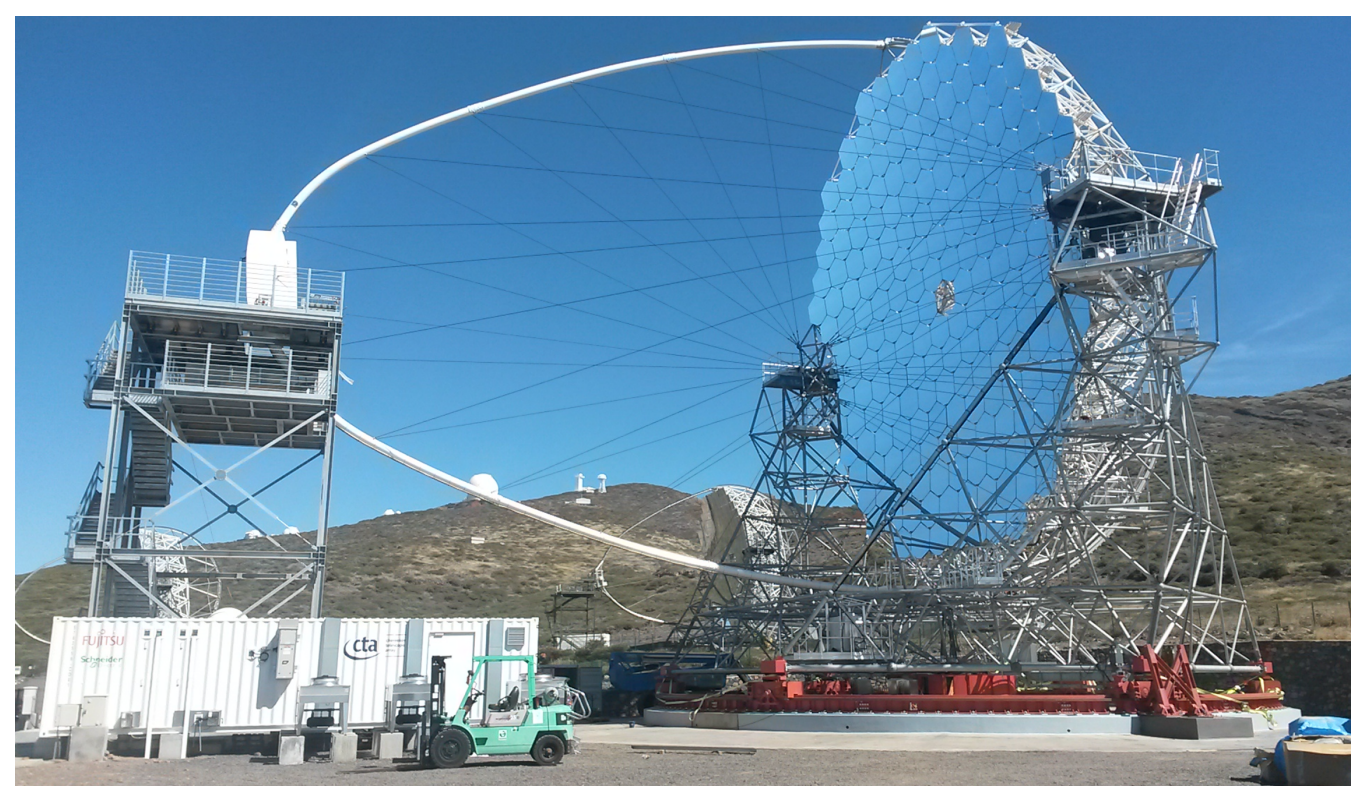

Figure 1. Picture of the 23m diameter LST at the Roque de los Muchachos observatory on the Canary island of La Palma (credits: LST collaboration)

However, the seismic requirements at the final CTA site in Chile have proven to be more demanding than originally anticipated. As a consequence, the current design would require additional reinforcement for two main reasons: first, to reduce the seismic response of the camera support structure and for the camera assembly; and second, to maintain sufficient preload in the Camera Support Structure (CSS) tethers, thereby preventing slackening during strong earthquake-induced vibrations.

In this article we describe our solution of a seismic isolation system including a seismic fuse that can be applied to the existing LST design with minor changes, but also it might inspire researchers to adopt a similar solution for their telescope designs.

## 2. LST-NORTH DESIGN

The overall LST structure is illustrated in Figure 2. The telescope employs a conventional altitude–azimuth (Alt-Az) configuration. The mechanical structure was designed and manufactured by MERO-TSK using its proprietary space-frame technology, which has been widely adopted in the construction of lightweight, large-span, and architecturally demanding structures. All structural components are certified and manufactured under stringent quality assurance and quality control procedures.

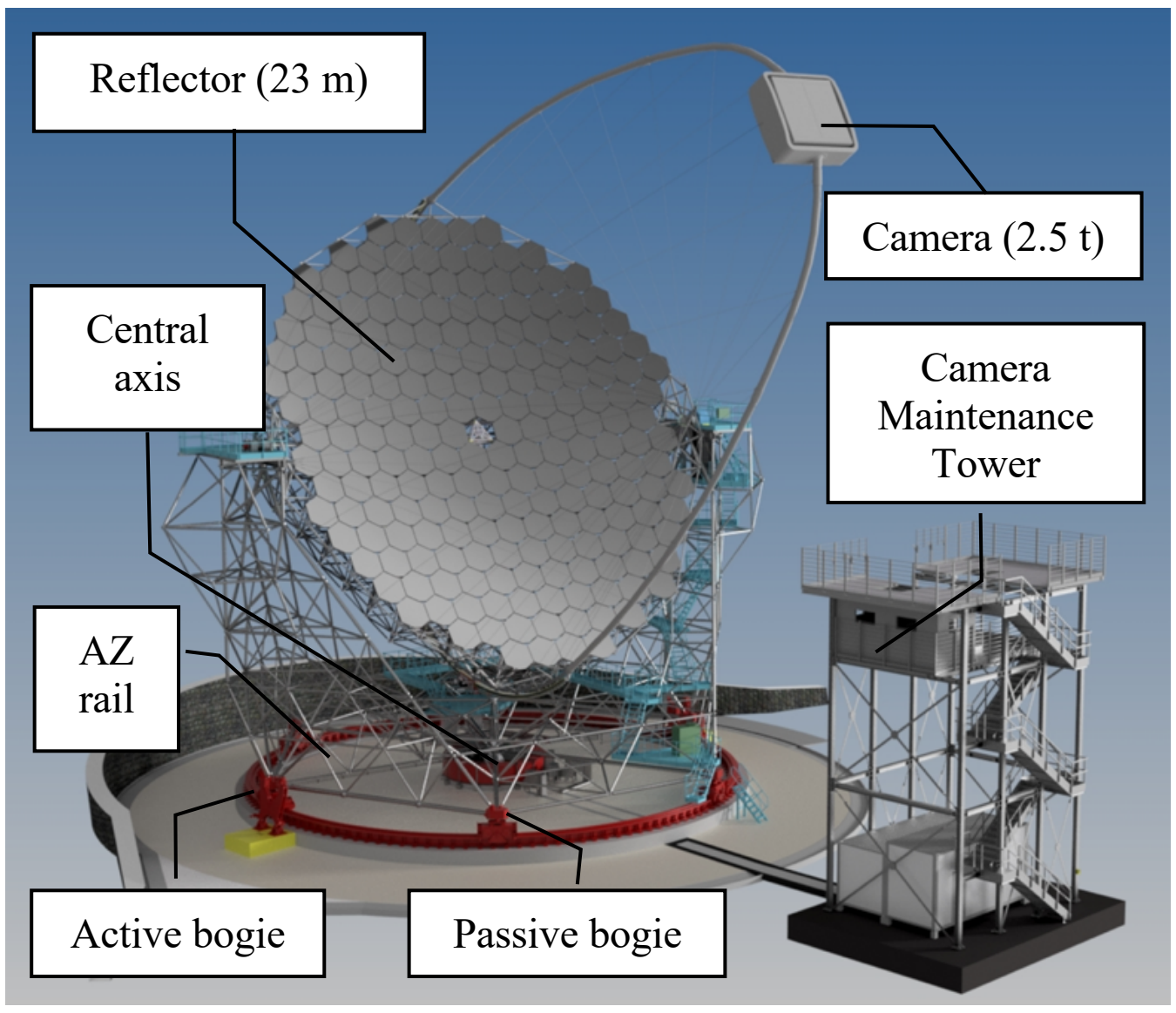

Figure 2. CAD model of the mechanical structure of the LST

The azimuth system is supported by four passive and two active bogies running on a circular rail. Both the rail and bogie assemblies were specifically engineered for the lightweight telescope structure and incorporate anti-uplift provisions to ensure structural stability under extreme wind loading conditions of up to 200 km/h. Due to the exceptionally low structural mass of the telescope relative to the aerodynamic wind forces, uplift effects become a critical design consideration. The reflector consists of 198 hexagonal aluminum–glass sandwich mirror panels equipped with active mirror control systems. The dish has a diameter of 23 m. Under the design survival wind speed of 200 km/h, the maximum quasi-static wind load reaches approximately 140 tons, exceeding the total mass of the moving telescope structure, which is only about 110 (in the abstract it is mentioned as 110 tons) tons.

The camera, with a mass of approximately 2.5 tons, is located 28 m from the dish vertex, coinciding with the elevation axis. It is supported by an ultra-lightweight camera support structure, manufactured from carbon-fiber-reinforced polymer (CFRP) and stabilized by a system of 26 CFRP tensioned ropes that have been preloaded between 2 and 5 tons. To minimize drive torques and enable rapid repositioning, the elevation assembly is counterbalanced by the rear support structure. The elevation drive system is directly coupled to the rear arch, providing efficient torque transfer and precise telescope positioning.

The following tables 1and 2 list the design parameters of the LST telescope. For structural sizing the wind gust (up to 140 tons), Vortex and the ice build-up (40 tons) are dimensioning (I have difficulties to understand the language of this sentence).

Table 1. Observation mode parameters

| **Observation mode slewing** | |
|---|---|
| • Azimuth angle range: | ±270 deg |
| • Elevation angle range: | +95 – 75 deg |
| • Azimuth drive max angular acceleration during repositioning: | 45.2 µrad/s$^2$ |
| • Azimuth drive max angular velocity during repositioning: | 8.4 mrad/s |
| • Elevation drive max angular acceleration during repositioning: | 0.7 µrad/s$^2$ |
| • Elevation drive max angular velocity during repositioning: | 72.6 µrad/s |
| **Observation mode emergency stop & GRB fast targeting** | |
| • Azimuth angle range: | ±270 deg |
| • Elevation angle range: | -70 deg to +90 deg |
| • Azimuth drive max angular deceleration: | 100 mrad/s$^2$ |
| • Elevation drive max angular deceleration: | 100 mrad/s$^2$ |

Table 2. Environmental parameters

| **Temperature** | |
|---|---|
| • Operational air temperature range: | -15°C to +25°C |
| • Air temperature change rate during night | <7.5 °C/h |
| • Survival air temperature range | -20°C to +40°C |
| **Wind** | |
| • Max. constant wind speed during observation | 36 km/h |
| • Max gust during observation | 60 km/h |
| • Max. constant wind speed during repositioning | 50 km/h |
| • Max gust during repositioning | 83.5 km/h |
| • Max. constant wind speed in safe state condition: 1. (in parking pos.) | 120 km/h |
| • Max. short term gust in safe state condition: (in parking position) | 200 km/h |
| **Ice and Snow** | |
| • Max. layer of snow in safe state condition | 500 mm |
| • Max. layer of ice in safe state condition: | 20 mm |

## 2. DIFFERENCE BETWEEN NORTH AND SOUTH SITES

The environmental conditions at the CTA North site on the Canary Island of La Palma and the CTA South site in Chile differ significantly. While the La Palma site is characterized by low temperatures, occasional ice accretion, and severe wind storms, it experiences only negligible seismic activity. In contrast, the Chilean site is located in a dry climate with generally less demanding wind conditions but is exposed to substantially higher seismic loads.

At the northern site, ice accumulation can occur under unfavorable combinations of humidity, temperature, and wind, resulting in thick ice deposits and asymmetric ice formations on structural components. Falling ice has occasionally caused damage to secondary infrastructure, such as stairs, platforms, and walkways. Owing to the significantly lower humidity levels at the Chilean site, such ice-related effects are not expected to occur.

From both a cost and a schedule perspective, the most efficient solution for the deployment of LSTs at the southern site would be the direct replication of the proven northern telescope design. The functional and scientific performance requirements for both sites are essentially identical; however, the environmental design loads differ substantially, particularly with respect to seismic loading conditions.

The following table summarizes the key environmental parameters that differ between the CTA North and CTA South sites, including the seismic design requirements specified for the southern site in Chile.

Table 3. Environmental parameters that differ between the two sites and seismic conditions

**Environmental parameters that differ between La Palma and Chile**

- Max. wind gust — 170 km/h in Chile vs 200 km/h in La Palma
- Ice layer — 20 mm of ice forming on all structure in La Palma
- Seismic activity — 0.43 pga horizontal in Chile vs 0.05 pga in La Palma

**Damage Limitation requirement (DLR)**

- Seismic action with a probability of exceedance P(DLR) = 10% in 10 years, TR=95 years respectively TR=475 in 30 years
- Structure must not yield and be back in operation after servicing

**No- Collapse requirement (NCR)**

- Seismic action with a probability of exceedance P(NCR) = 10% in 50 years, TR=475 years respectively TR=792 in 30 years
- Structure can suffer local permanent deformation but maintains its structural function and can be put back in operation at a reasonable cost.

The design response spectra adopted for the seismic qualification of all CTA telescopes at the Chilean site are presented in Figure 3. These spectra were derived specifically for the CTA South Observatory, taking into account site-specific seismic amplification effects associated with the local geotechnical conditions. The analyses consider variations in ground stiffness and stratigraphy, including the transition from competent bedrock to overlying soil and sedimentary layers, which can significantly influence the seismic response of the structures.

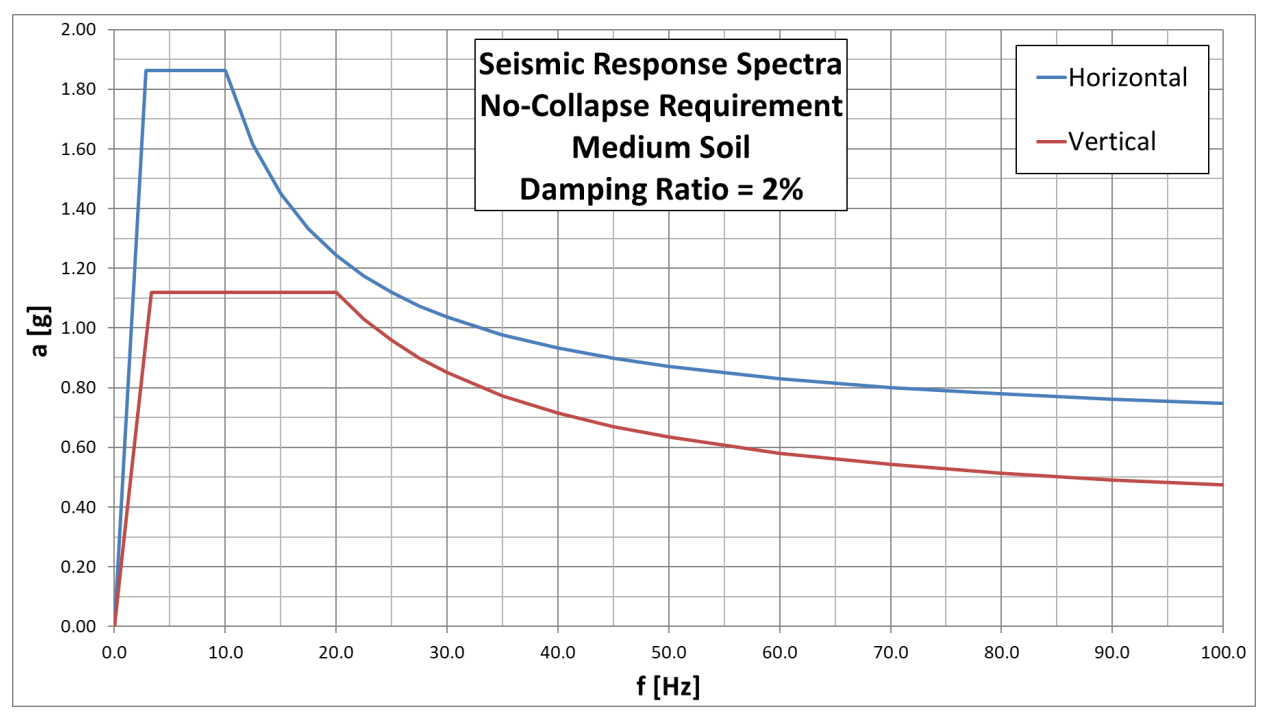

Figure 3. Seismic spectra defined by CTAO for the CTA Site in Chile. The fundamental frequencies of the telescope are at 1.9 Hz and higher (see Figure 8).

In order to check the compliance of the LST design with the environmental conditions at the south site, seismic response analyses have been performed with the result that non-negligible modifications would be needed at the camera support structure. In addition, the seismic response at camera level constitutes a high risk for the camera.

## 3. PROPOSED LST DESIGN FOR THE SOUTH SITE

The proposed concept combines the existing LST design with a stiff support structure equipped with a seismic isolation system, thereby reducing seismic dynamic amplification to acceptable levels. Since seismic mitigation measures are generally required for large telescope structures at the CTA South site, this approach enables the proven LST design to be retained largely unchanged while meeting the site-specific seismic performance requirements.

Several engineering approaches can be adopted to satisfy the seismic requirements of the CTA South site. The concept presented in this study focuses on seismic isolation as a means of protecting the existing LST structure. An alternative approach is to develop a telescope specifically optimized for the seismic environment of the site, as has been pursued for the new 23 m-class Cherenkov telescopes being developed by INAF. A further possibility would be to enhance the seismic resistance of the existing LST design through targeted structural reinforcement. Each of these approaches offers distinct advantages and represents a viable strategy for achieving the required seismic performance.

The following principal design requirements have been established:

- Ensure that LST-N telescopes can be used without significant changes, considering the DLR (no performance degradation) and NCR conditions (no non-recoverable performance degradation, no rupture of primary structure elements)
- Limit camera acceleration to 3.5g (goal)
- Ensure that CSS tether preloads are always in tension
- Provide a solid / stable basis for the telescope, especially during observation.
- Retain the telescope in aligned position up to wind speeds of 170km/h and up to a reasonable seismic level
- Release the telescope at a level ensuring damage limitation but above the equivalent wind loads at 170km/h.
- Facilitate re-alignment of the telescope position, in case of seismic levels above DLR level

### 4.1 Overall Concept

The proposed seismic isolation concept has been developed with the objective of minimizing modifications to the existing telescope design. To achieve this goal, the following measures are proposed:

• **Reinforcement of the azimuth rail support structure:** The annular rail support structure is strengthened to provide a rigid and stable support platform for the telescope. This is achieved by increasing the cross-sectional stiffness of the rail backup structure and by incorporating radial spokes to ensure high in-plane rigidity.

• **Distributed seismic isolation system:** A large number of seismic isolators are uniformly distributed beneath the rail support structure. The isolators are selected to provide high vertical stiffness while supporting the required horizontal flexibility.

• **Optimized isolator characteristics:** High-damping elastomeric bearings are employed, with their horizontal stiffness and damping properties tailored to the seismic isolation requirements. At the same time, the bearings provide high vertical stiffness to maintain accurate telescope alignment, as vertical isolation is not required.

• **Seismic fuse:** A replaceable seismic fuse is installed between the foundation and the rail support structure. The fuse is designed to release at a predefined load level, thereby limiting the forces transmitted to the telescope structure during extreme seismic events.

• **Post-earthquake realignment system:** Hydraulic jacks are incorporated to enable controlled repositioning and realignment of the telescope following a seismic event that has activated the seismic fuse.

Figure 4 illustrates the proposed seismic isolation concept together with the reinforced azimuth (AZ) rail structure. All modifications are confined to the region below the azimuth rail, allowing the telescope structure itself to remain essentially unchanged.

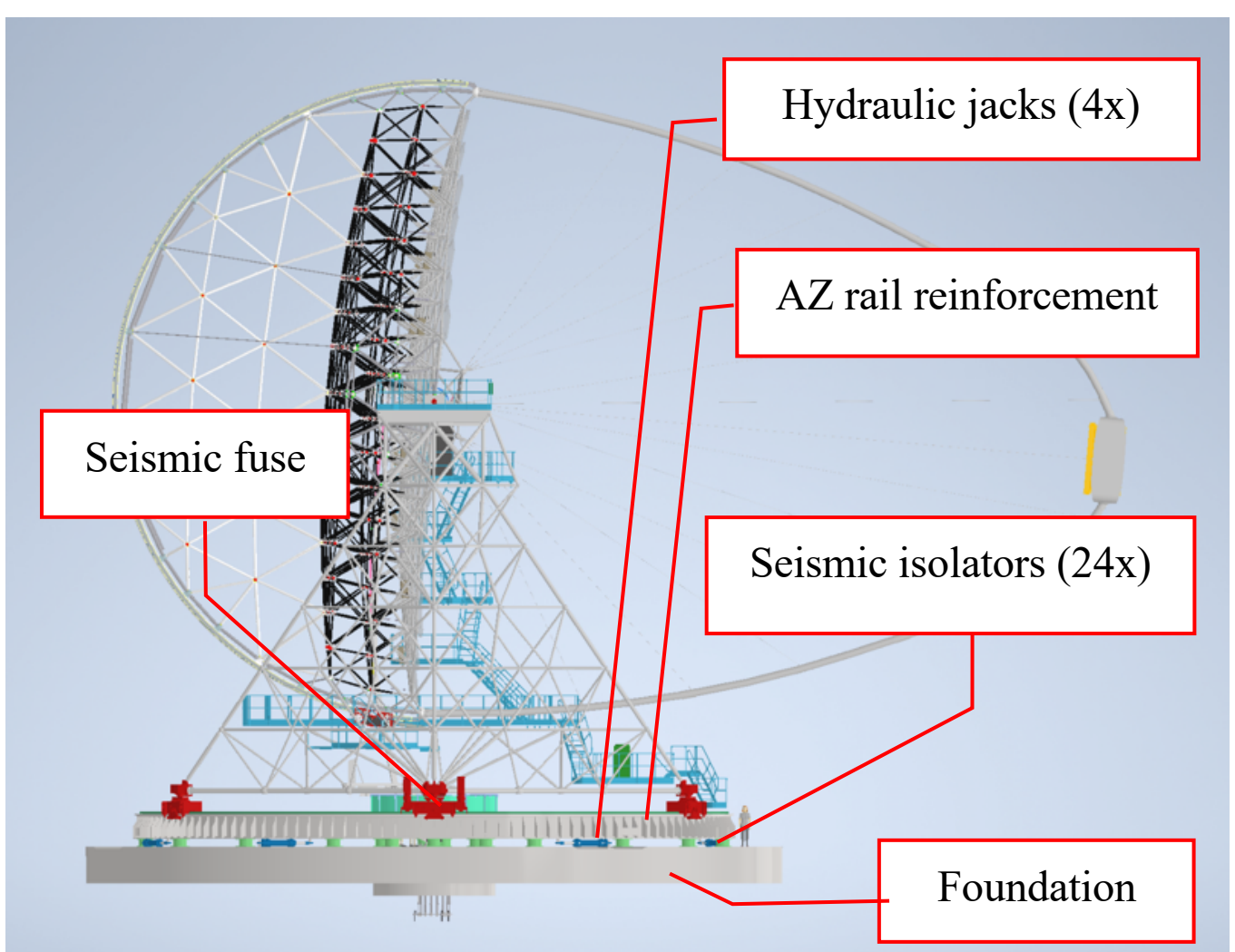

Figure 4: LST equipped with a seismic isolation system

### 4.2 Seismic isolation base plate and seismic fuse

Instead of constructing an additional reinforced concrete slab to provide a rigid support platform for the telescope, the azimuth rail structure is extended and reinforced below the bogie running surface. A series of radial stiffening members is incorporated, resulting in a wheel-like support structure with high in-plane stiffness. While the use of a dedicated concrete slab remains a viable alternative, the proposed reinforced rail concept offers an attractive and potentially more efficient solution.

The telescope axis is located at the centre of the spoke-wheel structure. The reinforced support wheel is connected to the foundation through a total of 24 seismic isolators, uniformly distributed along its inner and outer circumferences. The isolation system is designed to provide flexibility in the horizontal plane, thereby reducing the transmission of seismic forces to the telescope structure, while maintaining high vertical stiffness to support operational and environmental loads.

The centre of the spoke-wheel structure is connected to the foundation beneath the telescope's central azimuth bearing by means of a replaceable shear shaft. This shaft transfers the in-plane loads between the telescope and the foundation during normal operation and acts as a seismic fuse under extreme earthquake loading conditions. The fuse is designed to withstand all operational and survival wind loads while failing at a predefined seismic load level. This threshold is selected to remain below the load level that would cause permanent deformation of the telescope structure and below the level at which the Camera Support Structure (CSS) tethers could lose tension.

Following a major seismic event, the shear shaft can be replaced after the telescope has been repositioned to its nominal location. It is envisaged that this replacement procedure could be largely automated. After realignment of the telescope, a new shaft could be inserted through the mating lugs of the central rail hub and the foundation connection, thereby restoring the structural load path with minimal intervention and downtime.

The additional and modified components at the azimuth rail and foundation level are illustrated in the following figures.

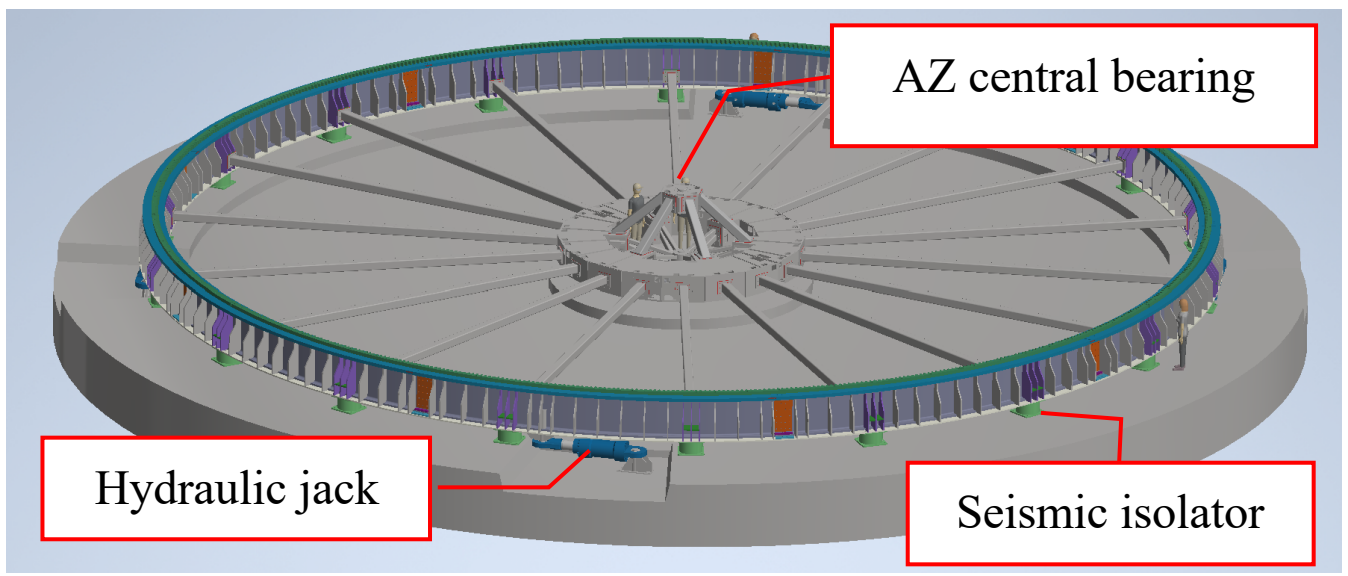

Figure 5: Seismic isolation system accommodated between rail and foundation

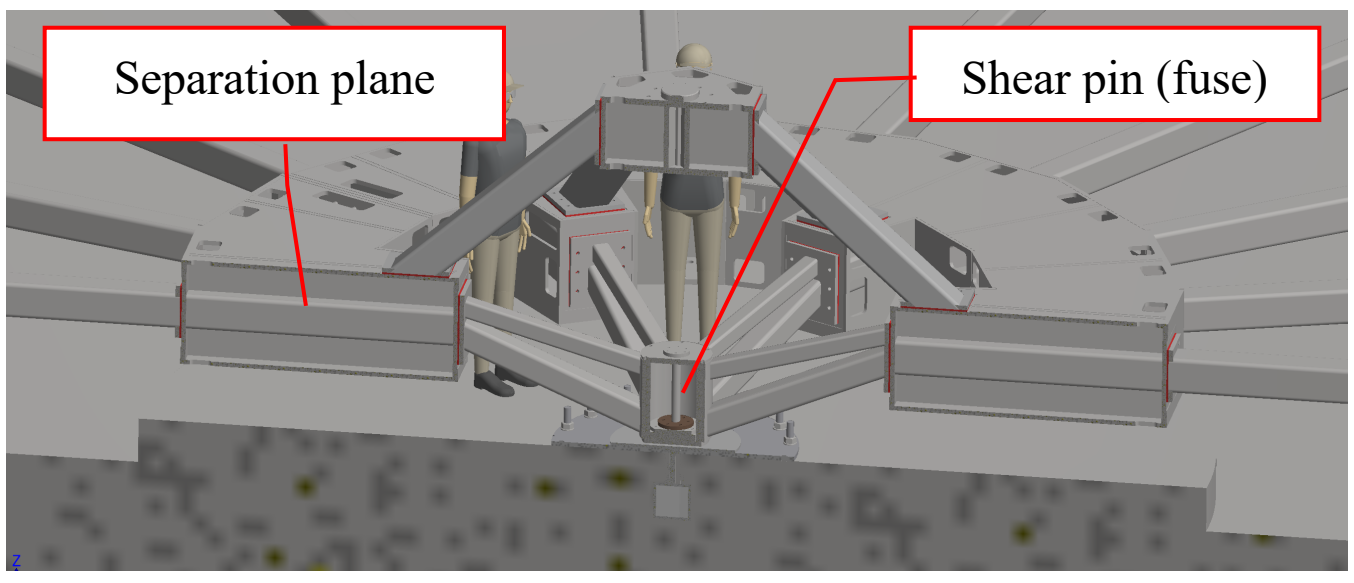

Figure 6: Accommodation of the seismic fuse below the AZ rotation axis

### 4.3 Commercially available seismic dampers

For the preliminary sizing of the seismic isolation system, high-damping rubber bearing (HDRB) isolators have been selected to reduce the seismic response of the telescope to acceptable levels. The objective is to ensure that the structure remains within the elastic regime under the specified operational earthquake level, thereby preserving functionality and limiting any post-event intervention to minor maintenance activities, while also preventing structural failure under the NCR-level earthquake.

The selected isolators consist of alternating layers of elastomer and steel laminates. The thickness and geometry of these layers are tailored to achieve the required horizontal stiffness and damping characteristics. Owing to their pronounced hysteretic behaviour, the isolators provide significant energy dissipation during seismic excitation, thereby reducing the dynamic response of the telescope. However, this hysteresis also limits the self-centering capability of the system, meaning that the telescope may not automatically return to its original position following a major seismic event.

The effectiveness of the seismic isolation system is governed primarily by the flexibility and damping characteristics of the isolators. The horizontal stiffness is selected to shift the fundamental frequency of the telescope-isolator system away from the dominant frequencies of the ground motion, thereby minimizing resonant amplification. At the same time, the inherent damping of the elastomeric bearings dissipates seismic energy and further reduces response amplitudes.

**4.4 Active repostioning by hydraulic actuators**

As a consequence of the limited self-centering capability of the isolators, active repositioning of the telescope may be required following a significant earthquake. This can be achieved by means of three hydraulic jacks arranged tangentially and uniformly distributed around the outer circumference of the spoke-wheel structure. The repositioning and relocking process can be fully automated through the use of position sensors, closed-loop control of the hydraulic actuators, and an automated replacement and insertion system for the central shear-fuse shaft.

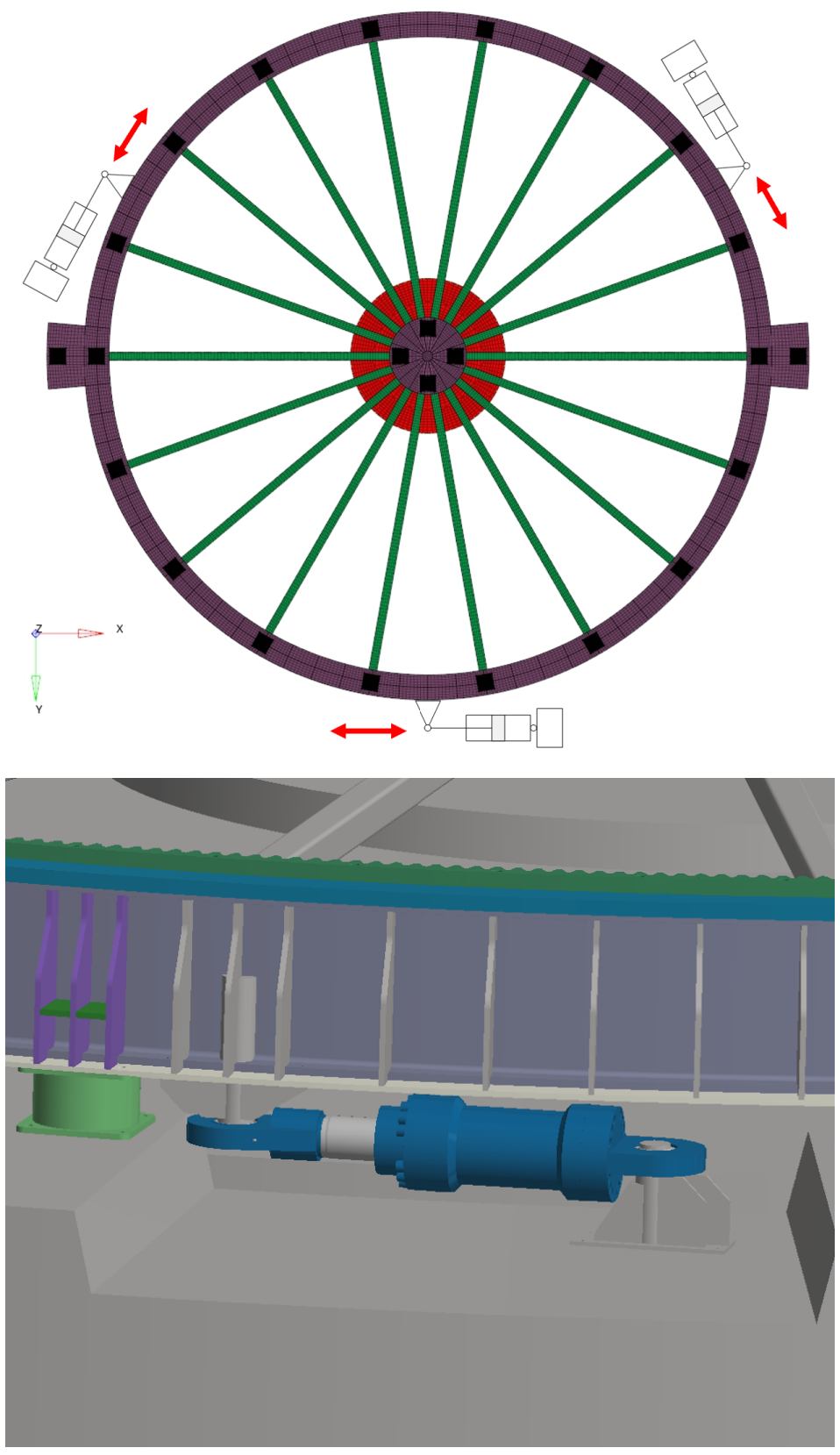

Figure 7: Accommodation of hydraulic jacks for repositioning of the telescope

An alternative seismic isolation concept based on hydraulic dampers in combination with curved-surface sliding bearings is also under consideration. Compared to the elastomeric bearing solution, this approach offers reduced mechanical complexity and may simplify installation and maintenance. However, additional analyses are required to assess its seismic performance, operational behaviour, and overall suitability for the LST application.

### 4.5 Seismic isolation requirements

Seismic response analyses were carried out to determine the required characteristics of the isolation system and to verify compliance of the telescope with the specified seismic performance requirements under the residual loading transmitted by the isolation system. The following parameters were identified as the governing design criteria:

• **Camera Support Structure (CSS) rope tension:** All CFRP tension members of the CSS shall remain under tension at all times. Rope slackening is not permitted under either static or dynamic loading conditions, including the design earthquake scenarios.

• **Camera acceleration:** The maximum acceleration at camera level shall not exceed 3.5 g.

Accordingly, the present study focuses primarily on these two governing performance parameters. Before evaluating the resulting structural response, the dynamic characteristics of the seismic isolation system are first assessed for both the locked and released configurations.

The following seismic performance states are defined:

• **Fuse release level:** The maximum earthquake level that can be sustained before activation of the seismic fuse. At this level, no structural damage or permanent deformation is permitted.

• **DLR-level earthquake with the isolation system released:** Following activation of the seismic fuse, the telescope shall withstand the DLR-level earthquake without permanent deformation or structural failure.

• **NCR-level earthquake with the isolation system released:** Following activation of the seismic fuse, the telescope shall withstand the NCR-level earthquake without rupture of any structural component.

## 5. FEM SIMULATIONS OF PROPOSED ISOLATION CONCEPT

The complete seismic isolation concept was modelled and analysed in detail, and the properties of the seismic isolators and associated mechanical parameters were optimized to assess both the benefits and the limitations of the proposed approach. The isolation system is designed to mitigate horizontal seismic forces only, while vertical seismic accelerations are transmitted directly to the telescope structure. This design choice represents a deliberate trade-off between isolation performance, system complexity, and implementation effort. By limiting the isolation functionality to the horizontal directions, the overall complexity and cost of the seismic isolation system can be significantly reduced while still achieving the primary objective of protecting the telescope from the most critical earthquake-induced loads.

### 5.1 Eigenmodes

The fundamental horizontal vibration modes of the telescope for both the locked and released configurations of the seismic isolation system are shown below. Upon activation of the isolation system, the fundamental natural frequency of the telescope is reduced from approximately 1.9 Hz to 1.0 Hz.

This frequency shift provides two important benefits. First, the fundamental mode is moved into a frequency range where the seismic input spectrum exhibits lower excitation levels, thereby reducing the dynamic response of the structure. Second, the isolation system attenuates the transmission of seismic energy from the ground to the telescope, reducing the dynamic coupling between the telescope structure, its subsystems, and the foundation.

The effectiveness of the seismic isolation system is governed primarily by the flexibility and damping characteristics of the isolators. Increased flexibility generally improves seismic isolation by lowering the fundamental frequency, although it must be balanced against practical limitations on allowable isolator displacement. Similarly, higher damping enhances energy dissipation and reduces response amplitudes but may increase residual displacements due to hysteretic behaviour, resulting in post-earthquake misalignment of the telescope.

The overall isolation performance can therefore be optimized through an appropriate selection of isolator stiffness, damping characteristics, and the number and distribution of isolators within the support system.

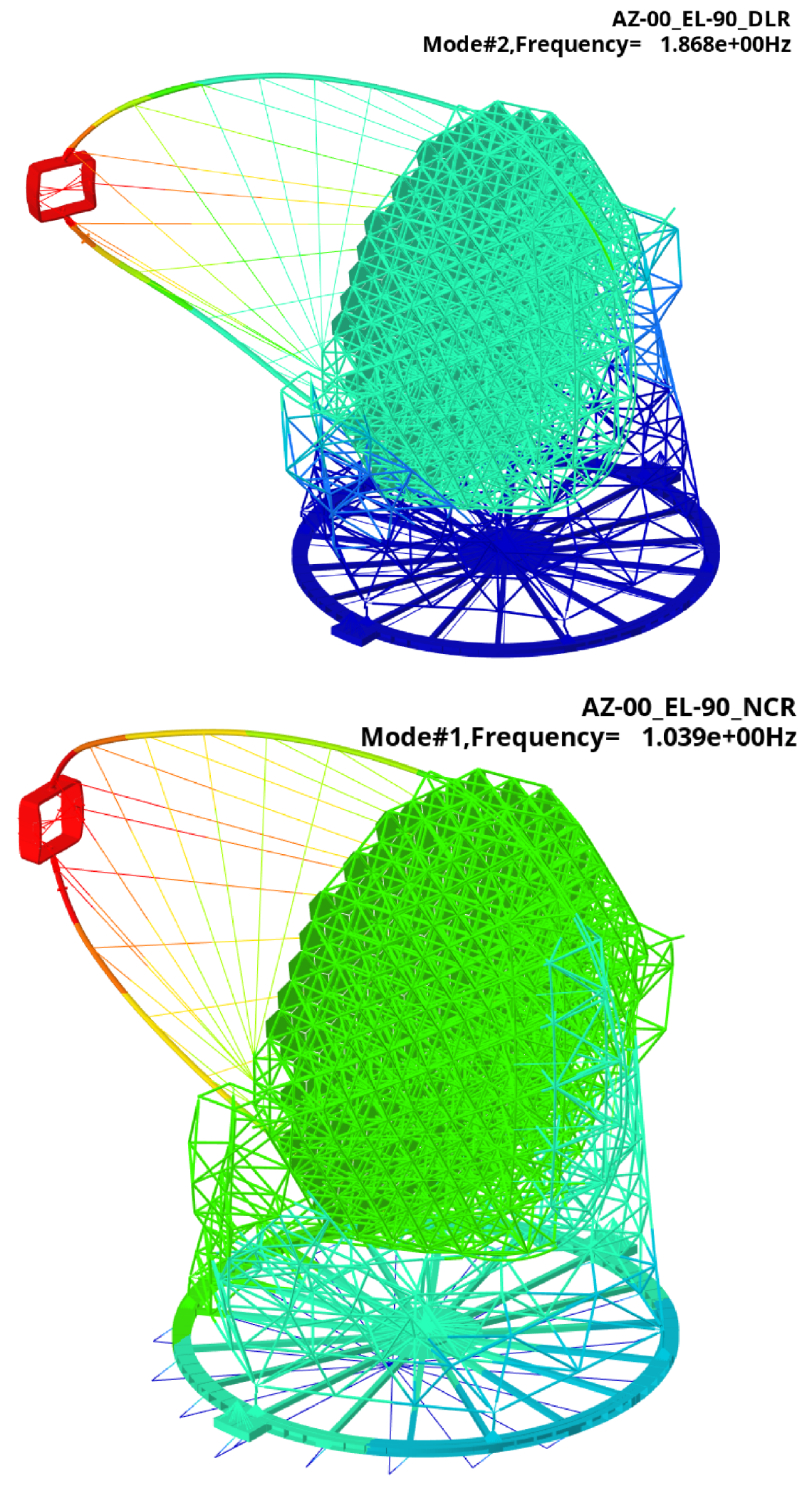

Figure 8: Comparison of eigenfrequencies in locked and released condition of the seismic isolation system.
Upper panel: Fused locked (f=1.9 Hz); Lower panel: Fuse released (f=1.04 Hz)

### 5.2 Load cases and input parameters for FEM simulation

The following load combinations were considered for the sizing and verification of the seismic isolation system:

- **G + P ± T ± W167**:
  Worst-case environmental loading condition
- **G + P ± T ± W83 ± FRL earthquake level**:
  Fuse release condition (isolation system locked; no yielding permitted)
- **G + P ± T ± W83 ± DLR-level earthquake**:
  Design-level earthquake with the seismic isolation system activated (no yielding permitted)
- **G + P ± T ± W83 ± NCR-level earthquake** :
  Maximum considered earthquake with the seismic isolation system activated (no structural rupture permitted)

where:

- **G**: Gravity load
- **P**: Preload in the Camera Support Structure (CSS) tethers
- **T**: Ambient temperature variation
- **W83**: Wind gust of 83.5 km/h (maximum operational wind gust)
- **W167**: Wind gust of 167 km/h
- **FRL**: Fuse Release Level
- **DLR**: Design-Level Requirement earthquake level
- **NCR**: No-Collapse Requirement earthquake level

Table 5. Environmental parameters that differ between the two sites and seismic conditions

| | Fuse release | DLR level | NCR level |
|---|---|---|---|
| Horizontal | 0.15 g | 0.25 g | 0.43 g |
| Vertical | N/A | 0.15 g | 0.26 g |

The fuse release level is selected such that there is no need for any relevant modification of the current design of the LST structure. The fuse release level shall be as high as possible to avoid frequent re-setting of the system and also to avoid release of the fuse under high wind speeds.

The following table 6 shows the selected design parameters and resulting conditions that have been used for the FEM analysis.

Table 6: Input parameters for FEM analysis

|  |  |
|---|---|
| **Mass** | |
| Telescope including bogies | 113 tons |
| Isolation structure including rails and central support | 114 tons |
| Total moving mass | 227 tons |
| **Isolator design parameters (preliminary nominal values prior to optimization based on analysis results)** | |
| Number of isolators | 24 |
| In-plane stiffness of isolators | 0.27 kN/mm |
| Out-of-plane stiffness of isolators | 469 kN/mm |
| Critical damping (not used in the response analysis) | 10% to 20% |
| **Displacement amplitude of isolators (horizontal)** | |
| Total out-of-plane stiffness | 6.48 kN/mm |
| Static amplitude under wind at 100 km/h (10 min average) | TBD mm |
| Dynamic amplitude under NCR level earthquake | TBD mm |
| | |
| **Reaction forces on fuse (fuse engaged)** | |
| Force on fuse from wind at 83.5 km/h (1s gust) | 309 kN |
| Force on fuse from wind at 100 km/h (10 min average) | 443 kN |
| Force on fuse from wind at 167 km/h (1s gust) | 862 kN |
| Force on fuse at release point | 995 kN |
| | |
| **Reaction forces on isolators (fuse released)** | |
| Force on isolators from wind at 83.5 km/h | 309 kN |
| Force on isolators from wind at 167 km/h | 1237 kN |
| Total horizontal force on isolators at NCR level | 1680 kN |
| Total vertical force on isolators at NCR level | 5080 kN |
| Max. horizontal force on a single actuator | 71 kN |
| Force needed for re-alignment | TBD kN |

The analyses were initially performed using a finite-element model representative of the current LST-N design, incorporating the proposed seismic isolation system, the modified azimuth rail structure, and a preliminary configuration of isolators with selected stiffness characteristics. Subsequently, an optimization process was carried out to ensure that the existing LST-N design can withstand the combined environmental and seismic loading conditions without requiring significant modifications to the telescope structure.

The analyses are conservative, as the enhanced damping provided by the high-damping isolators was not explicitly taken into account. This limitation arises because the analysis software does not allow separate damping properties to be assigned to the isolators and the telescope structure. Consequently, only the structural damping of the telescope was considered, resulting in a conservative estimate of the seismic response.

### 5.3 Structural Utilization Under Combined Static and Seismic Loading

The structural response of the telescope under combined static and seismic loading conditions, together with the resulting stress levels and utilization factors, is presented in the following sections. As the initial analyses did not demonstrate full compliance with the structural performance requirements, several design iterations were carried out to optimize the seismic isolation concept.

The development process proceeded as follows:

- **Baseline configuration without seismic isolation:** Initial analyses performed on the existing LST design without seismic isolation indicated excessive accelerations at camera level as well as elevated stress levels in several structural components. These results confirmed the need for a dedicated seismic isolation system.
- **Preliminary seismic isolation design:** A first isolation concept was developed, including an initial selection of the number of isolators and their stiffness characteristics. The analyses considered two operating conditions: (i) the seismic fuse engaged for earthquake levels up to the DLR level, and (ii) activation of the seismic isolation system following release of the fuse at earthquake levels exceeding the DLR threshold, with the maximum loading corresponding to the NCR-level earthquake.
- **Optimization of the fuse release level:** The initial configuration, with fuse release occurring at the DLR-level earthquake, still resulted in excessive utilization of the camera and the Camera Support Structure (CSS). To mitigate this issue, the fuse release threshold was reduced to approximately 60% of the DLR-level earthquake, thereby limiting the loads transmitted to the telescope while still in the locked configuration.
- **Optimization of isolator stiffness:** Although activation of the isolation system significantly reduced the seismic response, the camera and CSS continued to exhibit elevated stress levels under the NCR-level earthquake. Consequently, the horizontal stiffness of the isolators was further reduced to improve the isolation efficiency and achieve acceptable stress and utilization levels throughout the telescope structure.

The final results, presented in the following sections, demonstrate that all structural components satisfy the applicable design criteria and remain within the allowable stress and utilization limits.

### 5.4 FEM Simulation results and performance of the seismic isolation system

The simulations show encouraging results, basically reducing the seismic load on the telescope to such a degree that the camera and CFRP ropes are safe from damage or fatal destruction.

#### 5.4.1 Camera acceleration

The response acceleration of the camera is a good measure of the efficiency of the isolation system. The following results have been achieved.

- No isolation system 6.3 g
- Nominal isolation system 3.8 g
- Optimised isolation system 2.4 g at fuse release level, before release<br>3.5 g at released NCR level

Although no formal acceptance criterion has been specified for the camera, minimizing the loads transmitted to the camera remains a key design objective. Reducing camera stresses and accelerations provides an additional margin of safety and mitigates the risk of damage, which is particularly important given the substantial cost and downtime associated with camera repair or replacement.

### 5.4.2 Camera support structure (mast structure), camera frame and CFRP ropes

The camera support structure consists of the following components:

- **Camera frame and arch (both CFRP structures)**
- **CSS tether** (total of 26 ropes supporting the camera frame and arch)

The allowable loads for the ropes are:

- **GFRP ropes** 220 kN
- **Rope fixation: Permanent deformation** 65 kN
- **Rope fixation: Rupture** 90 kN

In addition, the CFRP ropes must remain under tension throughout the seismic event and must not become slack under dynamic loading. Loss of tension would introduce significant non-linear behaviour and could result in uncontrolled shock loading when the ropes transition abruptly from a slack condition back into tension. Such load reversals could generate local peak loads that are not captured by linear analysis methods and are therefore undesirable.

Since the ropes are already subjected to a substantial preload, it was considered preferable to lower the fuse release threshold to a level that prevents any loss of rope tension under the combined action of wind and seismic loading.

The resulting response for the load combination comprising an operational wind gust of 83 km/h and seismic loading at the fuse release level is summarized in table 7 below.

Table 7: Rope forces applied under static and seismic loads

| Rope forces | Static* | Dynamic* | Static+FRL | Static+NCR |
|---|---|---|---|---|
| Max | 44 kN | 23 kN | 66 kN | 78 kN |
| Min | 9 kN | 4 kN | -3 kN | -0.7 kN |

(*Where: Static G+P±W83 (gravity, preload, wind gust of 83km/h)
and Dynamic ±FRL (fuse release level) seismic

The strength requirements for the CFRP ropes are satisfied, although with only a limited margin. It can be observed that the NCR-level earthquake constitutes the governing load case and produces a more severe response than the fuse release condition. The DLR-level earthquake is fully enveloped by the NCR-level response, with the seismic isolation system activated in both cases.

While the fuse release threshold could be adjusted slightly, it should be noted that uncertainties must be taken into account. These include manufacturing tolerances affecting the actual rupture load of the seismic fuse, as well as variations in wind speed and wind direction that may occur simultaneously with the development of seismic ground motion. Consequently, a sufficient safety margin should be maintained when defining the final fuse release level.

It is to prevent the seismic fuse from release during a storm. The seismic fuse shall only be released during a powerful earthquake.

### 5.4.3 Reaction forces at the seismic fuse

Up to this point, the definition of the fuse release threshold has primarily been driven by seismic loading considerations. However, wind loading can also contribute significantly to the force acting on the fuse and may, under certain circumstances, trigger its release. Consequently, the selection of the fuse release threshold requires a careful compromise between seismic protection and avoidance of unnecessary fuse activations.

The most important parameters influencing both the nominal design release threshold and the actual release load are:

- The accuracy and repeatability of the fuse release mechanism (e.g., manufacturing tolerances and scatter in material properties)
- Telescope elevation angle, which influences both the aerodynamic loading and the dynamic response of the structure
- Wind speed and wind direction relative to the telescope azimuth angle
- Direction of the dominant seismic excitation relative to the telescope azimuth angle

The results summarized in the following table show that wind loading introduces a significant variation in the force acting on the fuse, depending on the instantaneous combination of telescope orientation, wind speed, and wind direction. A further important observation is that a front-wind condition at the maximum operational wind speed produces a fuse load exceeding the nominal release threshold. Under such circumstances, release of the fuse could be triggered by wind loading alone.

While such an event would not pose a risk to the structural integrity of the telescope, it would result in unnecessary operational downtime and require subsequent realignment of the telescope and replacement or re-engagement of the fuse.

It should be noted that the simultaneous occurrence of the most unfavorable combination of elevation angle, azimuth orientation, wind speed, wind direction, and seismic excitation direction represents a low-probability event. The actual loading state during an earthquake is therefore expected to be less severe than the envelope of all worst-case conditions considered independently.

The final selection of the fuse release threshold should be based on a more comprehensive parametric study. Such an assessment should also account for the fact that the specified wind speeds of 83.5 km/h and 167 km/h correspond to 1-second gusts, which may have a duration too short to fully activate the fuse release mechanism. Consequently, the fuse design and release mechanism should be calibrated to minimize the uncertainty band between the nominal and actual release thresholds while maintaining reliable and predictable operation. The following table 8 lists the resulting shear forces on the fuse.

Table 8: Shear force on the fuse under FRL earthquake and

| Direction | Elevation 0 deg | | Elevation 90 deg (parking) | | |
|---|---|---|---|---|---|
| | FRL | FRL+W83 | FRL | FRL+W83 | W167 kN |
| X | 494 kN | 494 kN | 477 kN | 477 kN | 0 kN |
| Y | 521 kN | 628 kN | 451 kN | 695 kN | 976 kN |
| Z | 0 kN | 0 kN | 0 kN | 0 | 0 kN |
| XY | 718 kN | 799 kN | 657 kN | 843 kN | 976 kN |

### 5.5 Seismic Rubber Isolators

The proposed concept is based on commercial seismic isolators, which are available in a broad range of sizes and stiffness characteristics. The selection of the most suitable isolator requires a detailed definition of the load, stiffness, damping, and displacement requirements derived from the present analyses.

#### 5.5.1 Isolator Properties

The forces acting on the isolators and the required displacement capacity derived from the analyses are summarized below and serve as the basis for the selection and sizing of suitable seismic isolators. In addition, the force acting on the seismic fuse is required for sizing the shear-pin diameter while accounting for uncertainties in both material properties and applied loading.

The isolator characteristics shall be selected to satisfy the force and stiffness requirements determined by the analyses. The values adopted in the present study are:

- **Vertical stiffness:** $\sim 0.469 \times 10^9$ N/m
- **Horizontal stiffness:** $\leq 0.27 \times 10^6$ N/m

The vertical stiffness is not a critical design parameter, provided that it is sufficiently high to maintain the required telescope alignment and pointing accuracy during operation.

The horizontal stiffness is a key parameter governing the seismic isolation performance and shall not exceed the specified value in order to achieve the required reduction in dynamic response. Based on the selected stiffness and the resulting wind and seismic load combinations, the following displacement capacities are required:

- **Required displacement capacity for the NCR-level earthquake:** $\geq 296$ mm
- **Resulting displacement under a 167 km/h wind load:** $\geq 76$ mm

The high damping required to minimize dynamic amplification is associated with pronounced hysteretic behaviour, as illustrated in the following figure. As a consequence, the telescope will not return automatically to its original position after a significant seismic event. Instead, it may come to rest at an arbitrary location within the displacement range of the isolation system, requiring subsequent realignment.

A possible realignment system based on hydraulic jacks is described in Section 4. The force required for repositioning is estimated to be approximately 40% of the maximum isolator reaction force.

The resulting design loads are:

- **Maximum shear force in a single isolator:** 88 kN
- **Maximum total isolator reaction force (sum of all isolator reactions):** 1,859 kN
- **Estimated force required for telescope realignment:** 744 kN

Appropriate design margins should be applied to account for uncertainties in loading, material properties, manufacturing tolerances, and long-term performance. Commercially available seismic isolators meeting these requirements are available from several manufacturers.

#### 5.5.2 Preliminary Isolator Requirements

Compared to the LST-N design, the proposed seismic isolation system introduces an additional rigid-body motion of the telescope at the azimuth rail level. The magnitude of this motion is governed primarily by the stiffness characteristics of the isolators and the seismic fuse. While the fuse limits horizontal translation of the isolation structure, the rotational motion (tilt) is controlled by the vertical stiffness of the isolators, which remain active and are not mechanically locked.

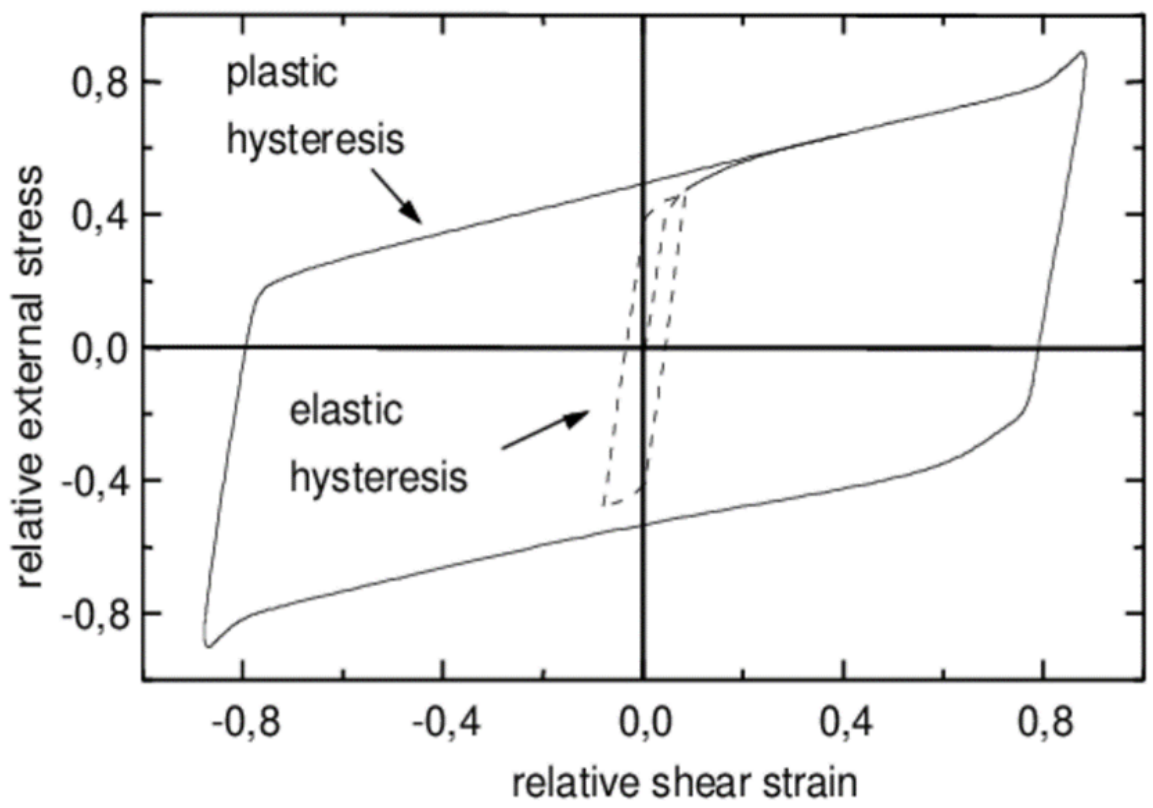

Figure 9: Typical hysteresis of an isolator

Figure 10 illustrates the rigid-body displacement of the telescope under a wind speed of 83.5 km/h. It should be noted that the operational requirements limit the maximum steady wind speed during observations to 36 km/h, with allowable wind gusts of up to 60 km/h. The corresponding rigid-body displacements of the isolation system are therefore significantly smaller and can be estimated as follows:

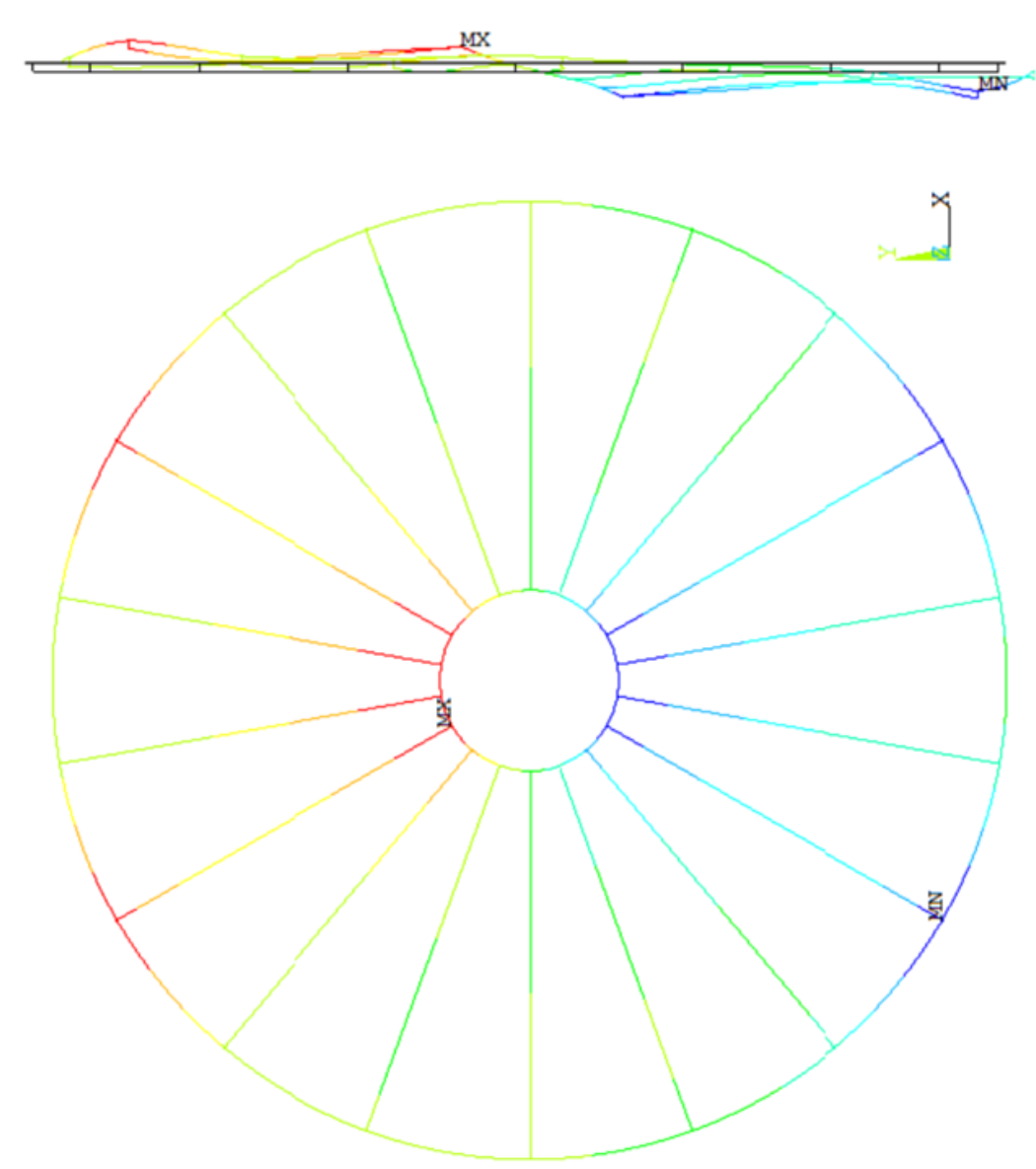

Upper panel: Side view Lower panel: top view

Figure 10: Isolation structure vertical displacement (red colour: max. 0.49 mm under wind of 83.5 km/h

**At the maximum steady operational wind speed of 36 km/h:**

- Horizontal displacement at rail level: 0.09 mm
- Tilt of the isolation structure: 0.031 mrad

**At the maximum operational wind gust of 60 km/h:**

• Horizontal displacement at rail level: 0.25 mm

• Tilt of the isolation structure: 0.085 mrad

These values represent only the rigid-body motion introduced by the seismic isolation system at the azimuth rail level. Structural deformations of the telescope itself must be superimposed on these displacements to obtain the total pointing error. However, the structural deformation component remains unchanged with respect to the LST-N design and is therefore identical to the values already established for the existing telescope. Corresponding deformation plots are presented in the following figures.

Table 8: Interface forces during seismic activity

| Interface | Direction | Unit | Allowable | Static | FRL | DLR | NCR |
|---|---|---|---|---|---|---|---|
| Passive bogies | Radial | kN | N/A | 0 | 0 | 0 | 0 |
| | Tangential | kN | N/A | 0 | 0 | 0 | 0 |
| | Vertical down | kN | 678 | **493** | 361 | 318 | 444 |
| | Vertical up | kN | -416 | **-251** | -126 | -94 | -235 |
| Active bogies/ALS | Radial | kN | N/A | 0 | 0 | 0 | 0 |
| | Tangential (ALS) | kN | 233 | 205 | 109 | 152 | **227** |
| | Vertical down | kN | 1145 | 422 | 719 | 677 | **881** |
| | Vertical up | kN | -517 | N/A | -143 | -148 | **-353** |
| Central pin | Lateral | kN | 901 | 109 | 432 | 430 | **661** |
| | Longitudinal | kN | | **598** | 324 | 329 | 468 |
| | Vertical | kN | 136 | 96 | **118** | 101 | 108 |
| CSS to dish | Lateral | kN | 58 | 0 | 1 | 1 | **1** |
| | Longitudinal | kN | | 37 | **49** | 44 | 48 |
| | Vertical | kN | 439 | 303 | **386** | 350 | 384 |
| | Bending | kNm | 27.8 | 11.4 | 22.1 | 17.5 | **22.7** |
| Elevation bearing | Radial | kN | 680 | 511 | 453 | 443 | **572** |
| | Axial | kN | 342 | 109 | 208 | 208 | **280** |
| Elevation drive | Direct | kN | 329 | 37.2 | **203** | 102 | 155 |
| CSS ropes | Tension | kN | 65/95 | 61.5 | 71.0 | 64.4 | **75.5** |

where

- Static Fuse locked G+P+W167
- FRL Fuse locked G+P+W+T+FRL
- DLR Fuse released G+P+W+T+DLR
- NCR Fuse released G+P+W+T+NCR

### 5.5.3 Preliminary Isolator Requirements

The interface forces acting on the sub-systems of the telescope are listed in the following table 8 together with the known allowable forces. The allowable loads are the maximum loads occurring at the north site. All forces from earthquakes are covered by the worst case conditions from the north site.

## 5.6 Camera Support Structure (CSS) Rope Loads

A detailed assessment was carried out for the prestressed CSS tension ropes. The following observations can be made:

- The rope preload fully compensates the effects of gravity and other static loads, including wind loading and temperature-induced deformations.
- Wind loading has only a minor influence on rope forces and becomes relevant primarily for northerly wind conditions when the telescope is in the parking position.
- Temperature effects are generally negligible. Only low-temperature conditions in the parking position produce a noticeable reduction in rope tension; however, this effect is adequately compensated by the initial rope preload.
- The combination of seismic loading with gravity and rope preload results in the most critical condition. In particular, the static-minus-seismic load combination leads to a slight exceedance of the allowable limits in the two central ropes. This behaviour is associated with the alternating nature of the seismic response.

The following figures illustrate the principal contributions to the rope forces for both zenith and horizon telescope configurations, considering the Fuse Release Level (FRL) earthquake with the fuse locked and the NCR-level earthquake with the fuse released. Seismic load combinations are generally evaluated together with an operational wind gust of 83.5 km/h, whereas the maximum wind speed of 167 km/h is considered only in combination with static loading in the parking position.

It can be observed that, for the NCR-level earthquake in the zenith configuration, a minor exceedance occurs in two of the twenty-six CSS tension ropes. All remaining ropes remain within the specified allowable limits.

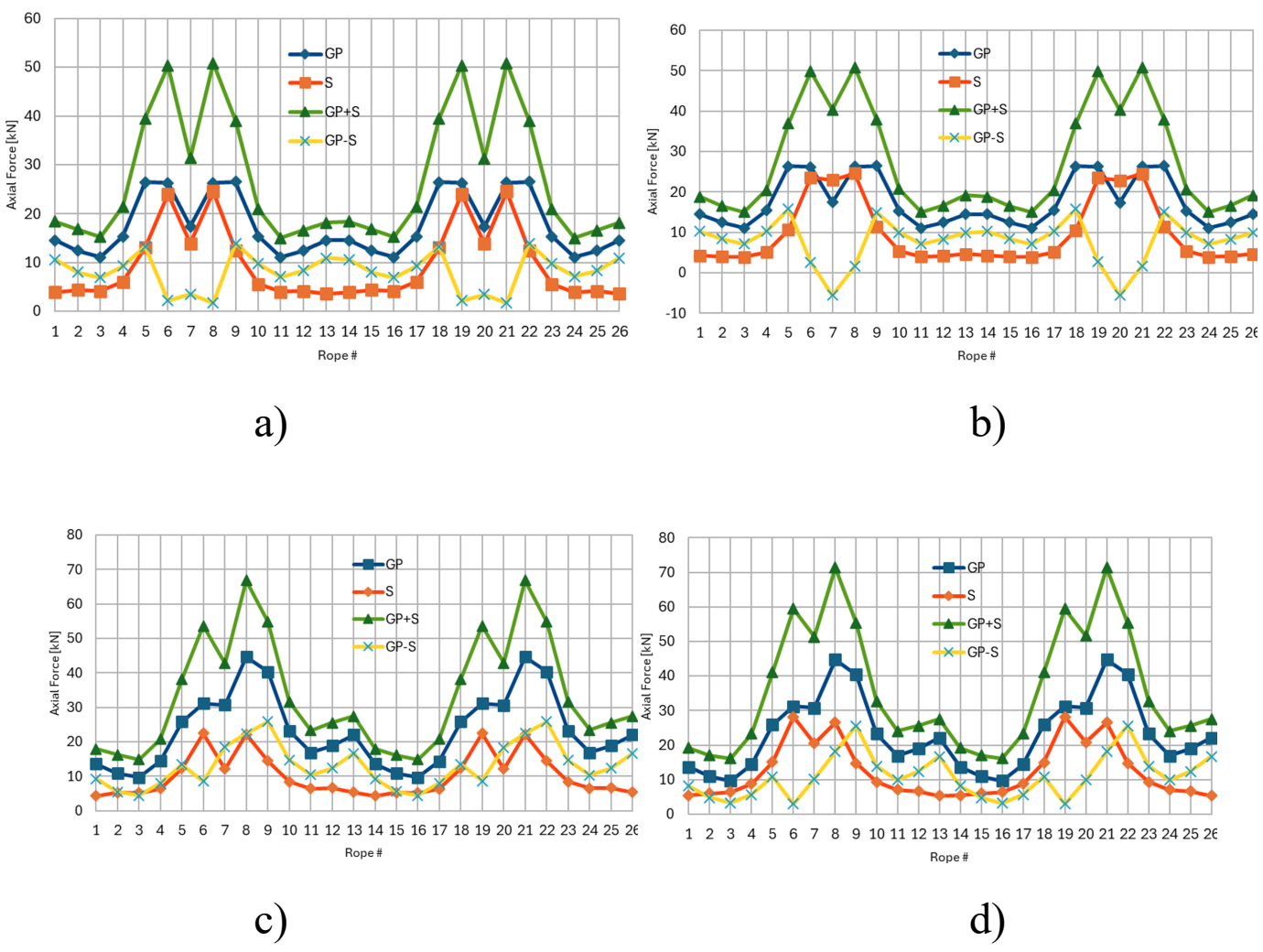

Figure 11: Rope forces from individual and combined load cases in Zenith configuration for a) FRL conditions, b) NCR conditions and in horizon configuration for c) FRL conditions, d) NCR conditions

### 5.7 Structural Load Capacity of the Telescope

Compliance of all primary structural components was demonstrated during the preliminary analyses, even for the initial, non-optimized seismic isolation configuration. Subsequent optimization of the isolation system further improved the structural response and increased the available safety margins. The final results are summarized in the following table.

The analyses considered the following load components and all relevant load combinations, as applicable for both the locked and released configurations of the seismic fuse:

- **G**: Gravity load (permanent)
- **P**: Preload in the prestressed CSS tension ropes (permanent)
- **T**: Maximum and minimum operating temperatures
- **W83**: Wind load corresponding to a wind speed of 83.5 km/h (combined with seismic loading to represent the most unfavorable operational condition)
- **W167**: Wind load corresponding to a wind speed of 167 km/h (combined with static loads G, P, and T to represent the most unfavorable environmental loading condition)
- **S**: Seismic loading (Fuse Release Level (FRL) earthquake for the locked-fuse configuration and NCR-level earthquake for the released-fuse configuration; the DLR-level earthquake is enveloped by the NCR-level response and is therefore not reported separately)

Stress, force, and utilization results are presented in the following tables and figures. Unless otherwise stated, the allowable values correspond to the onset of yielding for ductile components and to the ultimate failure load for non-ductile or rupture-critical components.

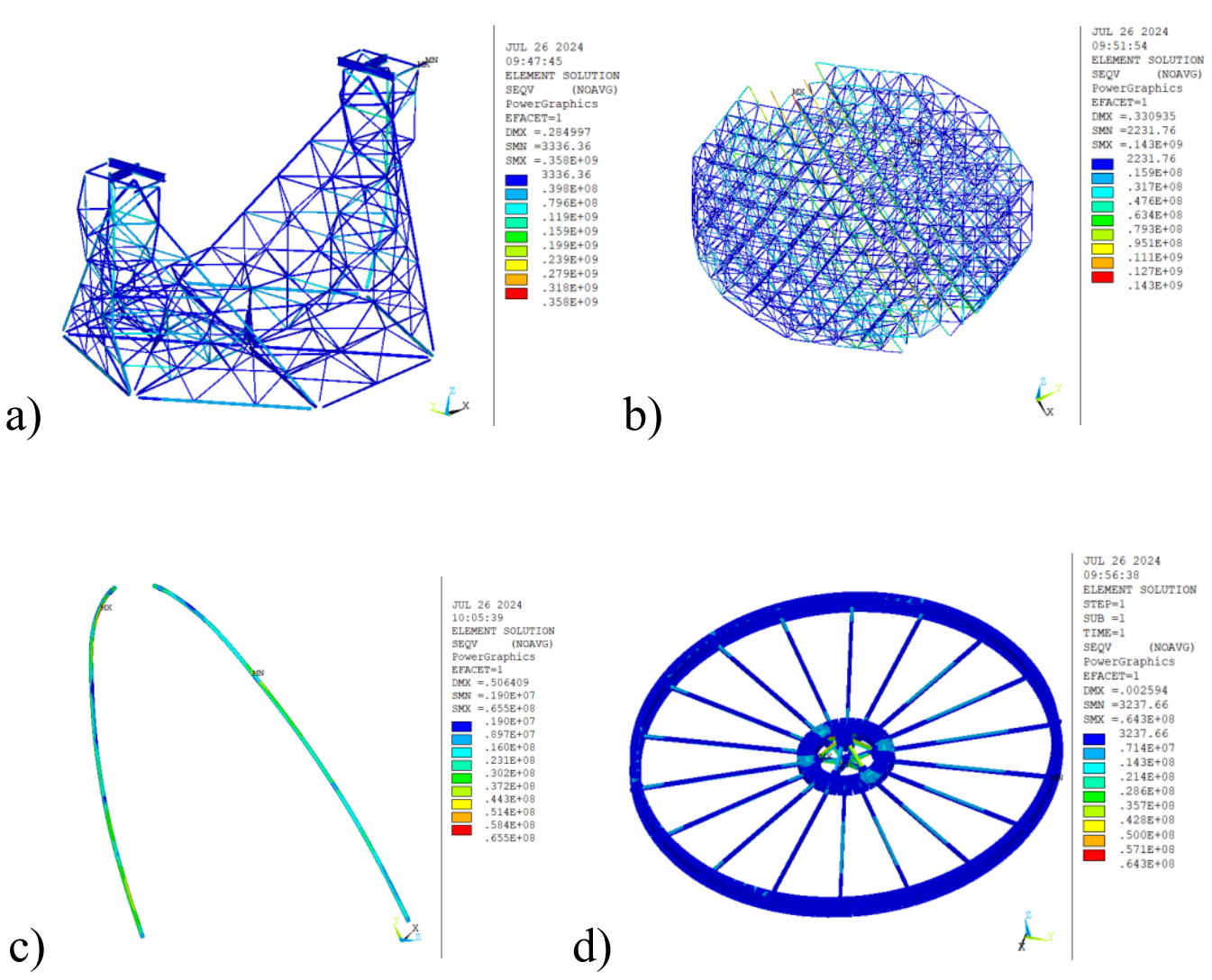

Figure 12: FEM Stress Distribution in the Main Structural Components
a) AZ structure, Zenith pointing, GP+S(NCR)+Wx83+T, equivalent stress (max. 358 MPa)
b) Dish CFRP structure, Zenith pointing, GP-S(NCR)+Wx83-T, equivalent stress (max. 143 MPa)
c) CSS CFRP structure, Zenith pointing, GP-S(NCR)+Wx83-T, equivalent stress (max. 66 MP
d) Isolation structure, Horizon pointing, GP+Wy167, equivalent stress (max. 64 MPa)

## CONCLUSION

The main conclusion is that, with the proposed seismic isolation system, the LST-N design can be made compliant with the LST-S requirements without significant modifications to the telescope itself. Two options are considered:

- Use the La Palma design with minimal modifications and mount the telescope on a large concrete slab connected to the foundation by a set of seismic isolators with the characteristics defined above.
- Reinforce the AZ rail cross-section and supplement it with a central hub and radial spokes to create a rigid support structure. The modified rail structure is then connected to the foundation via seismic isolators.

The performance does not depend on the material used for the support structure. Both solutions can be tuned to achieve the same performance, which is governed primarily by the fuse release level and the stiffness-to-mass ratio of the isolation system. The reinforced rail concept appears attractive, although the concrete slab solution may ultimately offer cost advantages.

The present analysis is based on a mechanical fuse combined with elastomeric dampers. However, a system based on curved-surface sliders combined with hydraulic dampers could potentially offer improved efficiency and reliability. Low-friction sliding bearings combined with hydraulic damping may also provide advantages in terms of self-centering and post-earthquake repositioning.

This study demonstrates the feasibility of the proposed concept and provides a conceptual design. More detailed analyses, combined with testing of scaled models, would be beneficial to further validate the proposed solution.

## ACKNOWLEDGEMENTS

Our seismic study is based on the mechanical structure of the existing LST at La Palma and we would like to acknowledge those who have participated the development of the seismic fuse and have played a significant role in the LST1 design. The LST is mostly designed, manufactured, assembled and commissioned by scientific institutes combined in the LST consortium with industrial contributions. We are proud of having achieved this great telescope and want to thank especially those involved in the telescope structure development, namely: Eckard Lorenz (MPP) for his great ideas and his guidance of the initial development and Klaus Charne (MERO-TSK) for the efficient collaboration, design and delivery of this special structure. Furthermore we would like to name the colleagues from the technical departments of the partner institutions for their pleasant and single-minded collaboration on interfaces: Laboratoire d'Annecy de Physique des Particules (LAPP), Annecy-le-Vieux, France: Dr. Armand Fiasson, Nicolas Geffroy, Guillaume Deleglise, Inocencio Monteiro, Institut de Física d'Altes Energies (IFAE), Barcelona, Spain : Dr. Oscar Blanch, Dr. Juan Cortina, Rafael Garcia, Juli Mundet, Universidad of Jaen, Spain: Prof. Antonio Joe Penuela Lopez, Maria Encarnacion Garrido Ruiz, Istituto Nazionale di Fisica Nucleare (INFN), Italy: Adriano Pepato. Also we would like to mention the colleagues from external companies supporting the analyses AIRWORKS: Alessandro Targusi, Enrico Sambenedetto, Martina Cocciolo, Luca Sillari, BURATTI: Enrico Buratti and GUIZZO SPACE: Gian Paolo Guizz We also acknlowledge our institute, Max-Planck-Institut für Physik in Garching, Germany who financed (supported ?) those studies.